\documentclass[prl,aps,letterpaper,floatfix,superscriptaddress,nofootinbib,preprintnumbers,twocolumn,10pt]{revtex4-1}

\usepackage{dcolumn}
\usepackage{bm}
\usepackage[dvips]{graphicx}
\usepackage{epsfig}
\usepackage{amsmath, amsfonts, amssymb, slashed}

\newcommand{\be}{\begin{equation}}
\newcommand{\ee}{\end{equation}}
\newcommand{\bee}{\begin{equation*}}
\newcommand{\eee}{\end{equation*}}

\usepackage{color}

\newcommand{\bea}{\begin{eqnarray}}  
\newcommand{\eea}{\end{eqnarray}}

\usepackage{feynmp}
\usepackage{tikz}
\begin{document}

\title{Is it $SU(2)_{\mathrm{L}}$ or just $U(1)_{\mathrm{Y}}$?\\ $750$ GeV di-photon probes of the electroweak nature of new states}

\author{Jose~Miguel~No}
\affiliation{Department of Physics and Astronomy, University of Sussex, Brighton 
BN1 9QH, United Kingdom}

\date{\today}

\begin{abstract}
The existence of a $750$ GeV di-photon spin-$0$ resonance $S$ would imply the additional presence of new particles beyond the Standard Model, coupling directly to $S$ 
and electromagnetically charged. For an $SU(2)_{\mathrm{L}}$ singlet $S$, 
we explore the possibility of probing the $SU(2)_{\mathrm{L}}$ and $U(1)_{\mathrm{Y}}$ quantum numbers of the new states at the LHC by measuring/constraining 
the $WW$, $Z\gamma$ and $ZZ$ decays of $S$. We obtain robust prospects on the required LHC integrated luminosity to discover the new decay modes of $S$, and discuss 
the implications of these measurements for probing the electroweak nature of the new states. 
We also discuss the impact of $S$ mixing with the SM Higgs in such probes. 
\end{abstract}

\maketitle

\section{I. Introduction}

\vspace{-3mm} 

Recently ATLAS~\cite{ATLAS_diphoton,ATLAS_diphoton_2}  and CMS~\cite{CMS:2015dxe,CMS:2016owr} collaborations have observed a large excess in the di-photon
spectrum around $m_{\gamma\gamma} \sim 750$ GeV, arising in LHC Run 2 data of {\it pp} collisions with center of mass energy 
$\sqrt{s} = 13$ TeV.
This excess strongly points towards the existence of a new neutral particle $S$ with mass $m_S \sim 750$ GeV, and such possibility has since been 
the subject of an intense research effort, particularly under the assumption of a spin-$0$ resonance. 

Being neutral, the new particle $S$ can only couple to photons via a loop of charged particles. These cannot be Standard Model (SM)
particles, since in such case the ratio of decay widths of $S$ into two such SM particles and into two photons  
would greatly suppress the di-photon decay mode, rendering it unobservable at the LHC (see {\it e.g.}~\cite{Knapen:2015dap,Ellis:2015oso}). 
This implies the existence of new electromagnetically charged particles $\chi$
beyond the SM, with masses $m_{\chi} \gtrsim 375$ GeV in order to forbid the decay of $S$ into these states. 
While the properties of these new $\chi$ particles may be constrained indirectly, {\it e.g.} via the measurement of running electroweak (EW) couplings at the 
LHC~\cite{Gross:2016ioi,Goertz:2016iwa}, a more direct probe would be ultimately required to establish their connection to the 750 GeV resonance 
phenomenology.

Due to $SU(2)_{\mathrm{L}} \times U(1)_{\mathrm{Y}}$ gauge invariance, the existence of the decay $S \to \gamma\gamma$ mediated by a loop of $\chi$ states 
automatically implies the existence of other decay modes~\cite{Low:2015qho}, namely $S \to Z\gamma$ and $S \to Z Z$, and also 
$S \to W W$ if $\chi$ transform non-trivially under $SU(2)_{\mathrm{L}}$.
Thus, measuring/constraining the various ratios of branching fractions $R_{VV}\equiv \mathrm{BR}(S \to V V)/\mathrm{BR}(S \to \gamma\gamma)$ (with $VV = WW,\,ZZ,\,Z\gamma$) 
could yield valuable information on the $SU(2)_{\mathrm{L}}$ and $U(1)_{\mathrm{Y}}$ quantum numbers of the new states $\chi$ mediating the decays 
of $S$. 

This paper is organized as follows: In Section II we discuss the relations between the di-photon signature and other potential signatures in the 
$WW$, $ZZ$ and $Z\gamma$ final states. In Section III we provide prospects for probing these accompanying final states during the 13 TeV Run of LHC in terms of 
required integrated luminosity $\mathcal{L}$. In Section IV we discuss the range of validity of our analysis, particularly regarding a possible mixing of 
$S$ with the SM Higgs. Finally, we conclude in Section V.

\vspace{-3mm} 

\section{II. $S$ decays beyond $\gamma\gamma$}

\vspace{-3mm} 

\begin{figure}[t!]
\centering
\includegraphics[width=0.485\textwidth]{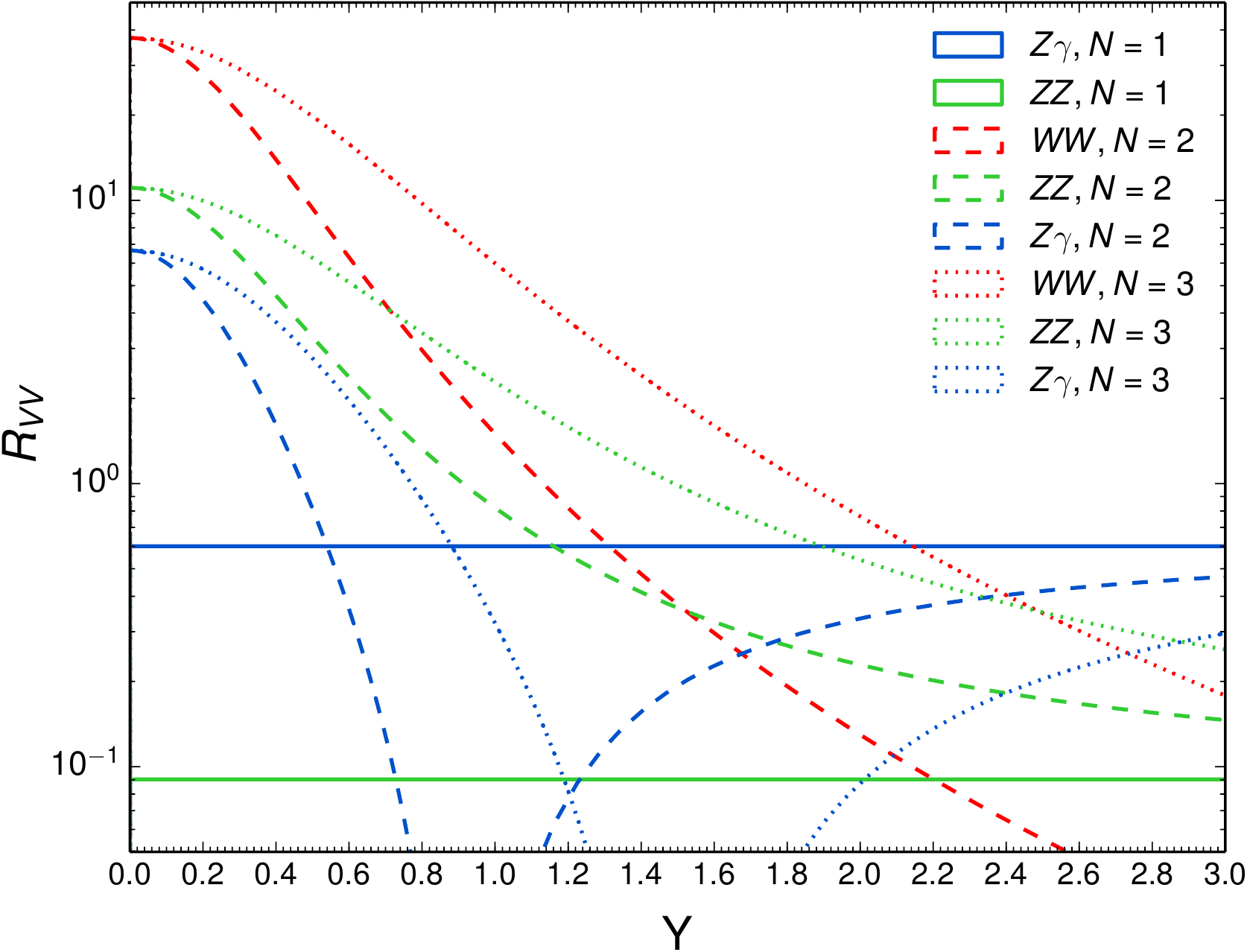}
\caption{$R_{VV}$ (with $VV = Z\gamma,\,ZZ,\,WW$) as a function of Y for $N = 1,\,2,\,3$ (for $N = 1$, Y $ > 0$ is implied).
}
\label{fig1:Ratio}
\end{figure}

In the following, I assume that the loop-induced decay $S \to \gamma\gamma$ is dominantly mediated 
by just one new particle species $\chi$ with certain EW quantum numbers, 
and that $S$ is (mostly) an EW singlet, such that potential tree-level decays of $S$ into $WW$ and $ZZ$ are suppressed compared to 
the corresponding loop-induced ones (I discuss the range of validity of this approximation in Section~IV). Then, the ratio 
of $S$-mediated di-photon production to that of $S$-mediated $WW$, $ZZ$ and $Z\gamma$ production is independent of the coupling $\lambda_{S\chi}$ between $S$ 
and $\chi$, its mass $m_{\chi}$ or the bosonic/fermionic nature of $\chi$: The ratios $R_{VV}$ depend solely on the EW quantum numbers of $\chi$, and as such 
they constitute an ideal observable for probing 
the $SU(2)_{\mathrm{L}} \times U(1)_{\mathrm{Y}}$ properties of $\chi$.
The various ratios $R_{VV}$ read, in the limit $m_S \gg m_W,m_Z$ (see {\it e.g.}~\cite{Low:2015qho,Franceschini:2016gxv,Altmannshofer:2015xfo})
\begin{eqnarray}
\label{ratio1}
R_{Z\gamma} & =& \frac{2\, (\frac{c_W}{s_W} - \kappa \frac{s_W}{c_W})^2}{(1+\kappa)^2}\\
\label{ratio2}
R_{ZZ} & =& \frac{(\frac{c^2_W}{s^2_W} + \kappa \frac{s^2_W}{c^2_W})^2}{(1+\kappa)^2}  \\ 
\label{ratio3}
R_{WW} & =& \frac{ 2}{s^4_W\,(1+\kappa)^2} 
\end{eqnarray}
with $s_W$ ($c_W$) being the sine (cosine) of the Weinberg angle. As~\eqref{ratio1}-\eqref{ratio3} show, all three ratios $R_{VV}$ are fully 
controlled by the one parameter $\kappa$ if $S$ is an un-mixed $SU(2)_{\mathrm{L}} \times U(1)_{\mathrm{Y}}$ gauge 
singlet\footnote[1]{In contrast, if $S$ contains an admixture of the SM Higgs doublet (or another new scalar, transforming non-trivially 
under $SU(2)_{\mathrm{L}}$~\cite{Howe:2016mfq}), then tree-level decays into $WW$ and $ZZ$ introduce an extra parameter dependence~\cite{Franceschini:2016gxv}.}.
The value of $\kappa$ is given by
\begin{equation}
\label{kappa}
\kappa = \frac{12\,\mathrm{Y}^2}{(N-1)(N+1)} 
\end{equation}
where Y is the hypercharge of $\chi$ and $N = 1,\,2,\,3,\,4...$ denotes its $SU(2)_{\mathrm{L}}$ representation
($1 =$ singlet, $2 =$ doublet, $3 =$ triplet...). In Figure~\ref{fig1:Ratio} we show the values of $R_{VV}$ as a function 
of Y for $N = 1,\,2,\,3$. The values of all $R_{Z\gamma}$, $R_{ZZ}$ and $R_{WW}$ are maximized for $\mathrm{Y} = 0$ ($\kappa = 0$, 
with $\chi$ being a pure $SU(2)_{\mathrm{L}}$ state), while for $\chi$ being a pure $U(1)_{\mathrm{Y}}$ state  
($N=1$, $\kappa = \infty$), both $R_{ZZ}$ and $R_{WW}$ reach their minimum values $R_{ZZ} = 0.0902$ and $R_{WW} = 0$, while 
$R_{Z\gamma} = 0.6008$ (the ratios are independent of the value of Y, but Y $ > 0$ is implied). 
As seen from~\eqref{ratio1}, $R_{Z\gamma}$ reaches its minimum $R_{Z\gamma} = 0$ for $N >1$ and $\mathrm{Y}^2 = c^2_W(N-1)(N+1)/(12\,s^2_W)$. 
We also note that while for $\kappa = 0$ we have $R_{WW} > R_{ZZ} > 1$, as $\kappa$ increases there is a turning point 
above which $1 > R_{ZZ} > R_{WW}$, which occurs for $\kappa > c^2_W (\sqrt{2} - c^2_W)/s^4_W$.

We stress that, as is apparent from the relations~\eqref{ratio1}-\eqref{kappa}, for $N > 1$ only a combination of $N$ and Y can be accessed by the 
measurement of the ratios $R_{VV}$, but not the $SU(2)_{\mathrm{L}}$ representation of $\chi$ alone. However, such measurement of the ratios does provide 
disentangling power among $N = 1$,  $N > 1$ with $\mathrm{Y}>0$, and $N > 1$ with $\mathrm{Y}=0$. In the next Section we analyze the prospects for LHC 13 TeV
(as well as the constraints from LHC 8 TeV) in this respect.

\vspace{-3mm} 
 
\section{III. Prospects for LHC 13 $\mathrm{TeV}$}
 
\vspace{-3mm}  

In order to analyze the LHC 13 TeV prospects of probing the di-photon resonance $S$ accompanying decays $S \to Z \gamma$, $S \to Z Z$ and $S \to W W$, 
and provide a robust estimate of the integrated luminosity $\mathcal{L}$ needed to exclude at 95\% C.L. the presence of 
these new decay modes for different $SU(2)_{\mathrm{L}}$ representations and values of the $U(1)_{\mathrm{Y}}$ hypercharge Y, 
the di-photon cross section $\sigma(p p \to S \to \gamma\gamma)$ required to fit 
the ATLAS and CMS 13 TeV data is needed. Under the assumption of a narrow width\footnote[2]{The criterium for NW {\it vs} LW refers to experimental mass resolution. 
We note that for LW with $\Gamma_S \sim 45$ GeV, the ratio $\Gamma_S/m_S \sim 0.05$, so that $S$ may still be considered a narrow resonance in the broad sense.} (NW)
for $S$, the central value (CV) and the lowest value allowed at 95\% C.L. for the di-photon cross section from a fit to the 13 TeV ATLAS $3.2\,\mathrm{fb}^{-1}$ of
data~\cite{Buckley:2016mbr} are respectively given by
\begin{equation}
\label{ATLAS_NW}
\sigma^{\mathrm{NW-CV}}_{\gamma\gamma_{\mathrm{ATLAS}}}  \simeq 7.5 \,\, \mathrm{fb} \quad ,\quad 
\sigma^{\mathrm{NW-95\%}}_{\gamma\gamma_{\mathrm{ATLAS}}}  \simeq 3.1 \,\, \mathrm{fb}\,.
\end{equation}
Similarly, the respective di-photon cross sections from the 13 TeV CMS di-photon $3.3\,\mathrm{fb}^{-1}$ data fit~\cite{CMS:2016owr} are
\begin{equation}
\label{CMS_NW}
\sigma^{\mathrm{NW-CV}}_{\gamma\gamma_{\mathrm{CMS}}}  = 4.87 \,\, \mathrm{fb} \quad \, ,\quad 
\sigma^{\mathrm{NW-95\%}}_{\gamma\gamma_{\mathrm{CMS}}}  = 0.91 \,\, \mathrm{fb}\,.
\end{equation}
In contrast, for a wide resonance$^2$ (LW)
, with $\Gamma_S \sim 45$ GeV, the above di-photon cross sections read 
\begin{equation}
\label{ATLAS_LW}
\sigma^{\mathrm{LW-CV}}_{\gamma\gamma_{\mathrm{ATLAS}}}  \simeq 13 \,\, \mathrm{fb} \quad, \quad 
\sigma^{\mathrm{LW-95\%}}_{\gamma\gamma_{\mathrm{ATLAS}}}  \simeq 6.5 \,\, \mathrm{fb}\,,
\end{equation}
\begin{equation}
\label{CMS_LW}
\sigma^{\mathrm{LW-CV}}_{\gamma\gamma_{\mathrm{CMS}}}  \simeq 4.2 \,\, \mathrm{fb} \quad, \quad 
\sigma^{\mathrm{LW-95\%}}_{\gamma\gamma_{\mathrm{CMS}}}  \simeq 0.8 \,\, \mathrm{fb}\,.
\end{equation}
The ATLAS cross sections in \eqref{ATLAS_LW} are again obtained from~\cite{Buckley:2016mbr}, while the CMS ones 
are obtained via a proper rescaling of the NW cross sections in \eqref{CMS_NW} using 
the respective significances from the $m_{\gamma\gamma} = 750$ GeV $p$-values for NW ($\Gamma_S/m_S = 0.00014$) and LW ($\Gamma_S/m_S = 0.056$) from~\cite{CMS:2016owr}.  
We remark that our choice of di-photon signal cross section benchmarks (CV and lowest value allowed at 95\% C.L.) is motivated by the fact that significantly higher values 
(in particular the highest allowed value at 95\% C.L. for $\sigma_{\gamma\gamma}$ from 13 TeV LHC data) are excluded at more than 
95\% C.L. by LHC 8 TeV CMS di-photon data~\cite{Khachatryan:2015qba}. 

Both ATLAS and CMS have very recently performed searches for heavy spin-$0$ resonances decaying to $Z\gamma$~\cite{ATLAS13Zgamma,CMS:2016rsl}, $ZZ$~\cite{ATLAS13ZZ,CMS:2016noo} 
and $WW$~\cite{ATLAS13WW} at 13 TeV,
providing 95\% C.L. upper bounds on the respective cross sections. 
Focusing on the most stringent limits for each channel\footnote[3]{
The ATLAS bounds are obtained with $3.2\,\mathrm{fb}^{-1}$ of data, {\it vs} the CMS $2.7\,\mathrm{fb}^{-1}$ 
($2.3\,\mathrm{fb}^{-1}$) of data in the $Z\gamma$ ($ZZ$) analysis. The ATLAS constraints are generically stronger, with the sole exception of 
the LW scenario for $Z\gamma$ and $ZZ$, for which an ATLAS analysis does not exist, while a CMS one does.}, in combination with  
the required range of di-photon cross sections \eqref{ATLAS_NW}-\eqref{CMS_LW} these allow to constrain the value of the various ratios $R_{VV}$ in \eqref{ratio1}.
Since the improvement on the 13 TeV bounds is expected to scale 
simply as $\sim\sqrt{\mathcal{L}}$ in the future (as the current bounds are already obtained from 13 TeV data, the signal and background efficiencies of these analyses are expected 
not to change significantly as more data is collected), it is then possible to obtain a fairly robust estimate of the required amount of integrated luminosity 
$\mathcal{L}$ to exclude at 95\% C.L. the presence of the decay modes $S \to Z\gamma,\,ZZ,\,WW$ for different $SU(2)_{\mathrm{L}}$ representations and 
values of the $U(1)_{\mathrm{Y}}$ hypercharge Y. As discussed at the end of Section~II, there are three qualitatively distinct scenarios to be probed: The pure 
$U(1)_{\mathrm{Y}}$ scenario, the pure $SU(2)_{\mathrm{L}}$ scenario and the case of $\chi$ being charged under both. These three scenarios can be fully explored 
then by considering the case of $\chi$ as an $SU(2)_{\mathrm{L}}$ singlet with Y $ > 0$ and the case of $\chi$ as an $SU(2)_{\mathrm{L}}$ doublet with Y $ \geq 0$.
%
%
%

\begin{widetext}
\vspace{-1mm}
\onecolumngrid\begin{figure}[h!]
\centering
\includegraphics[width=0.49\textwidth]{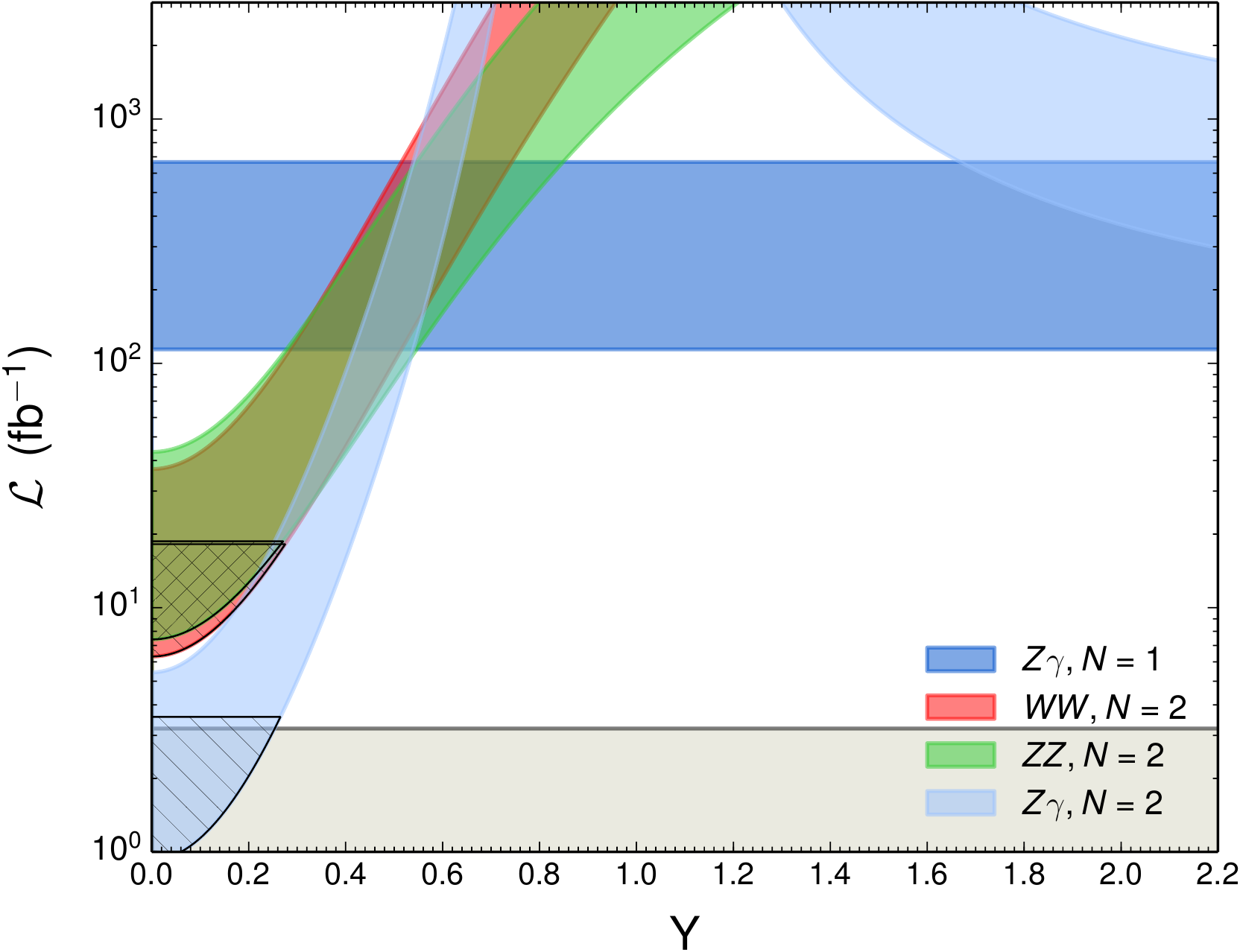}
\includegraphics[width=0.49\textwidth]{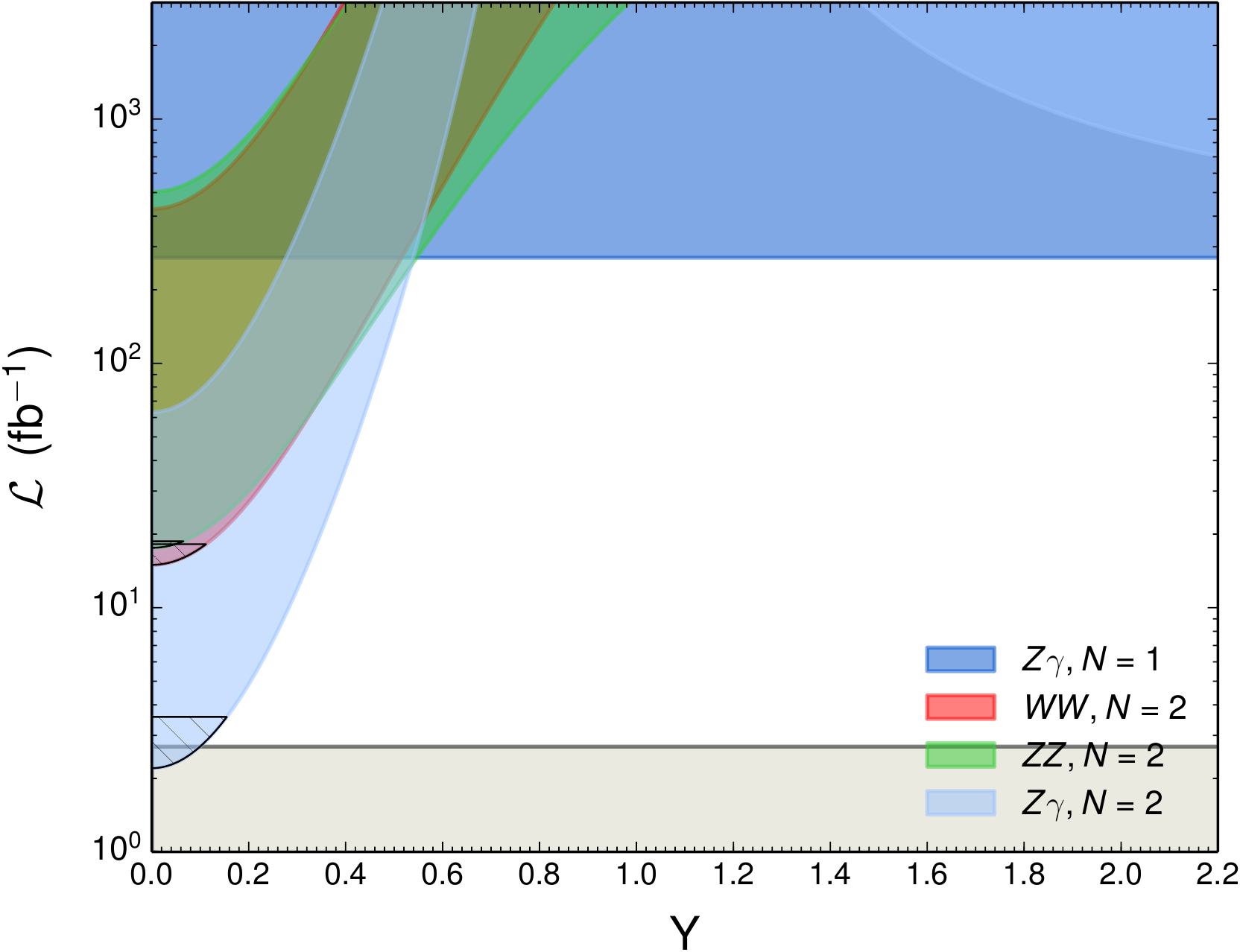}

\vspace{2mm}

\includegraphics[width=0.49\textwidth]{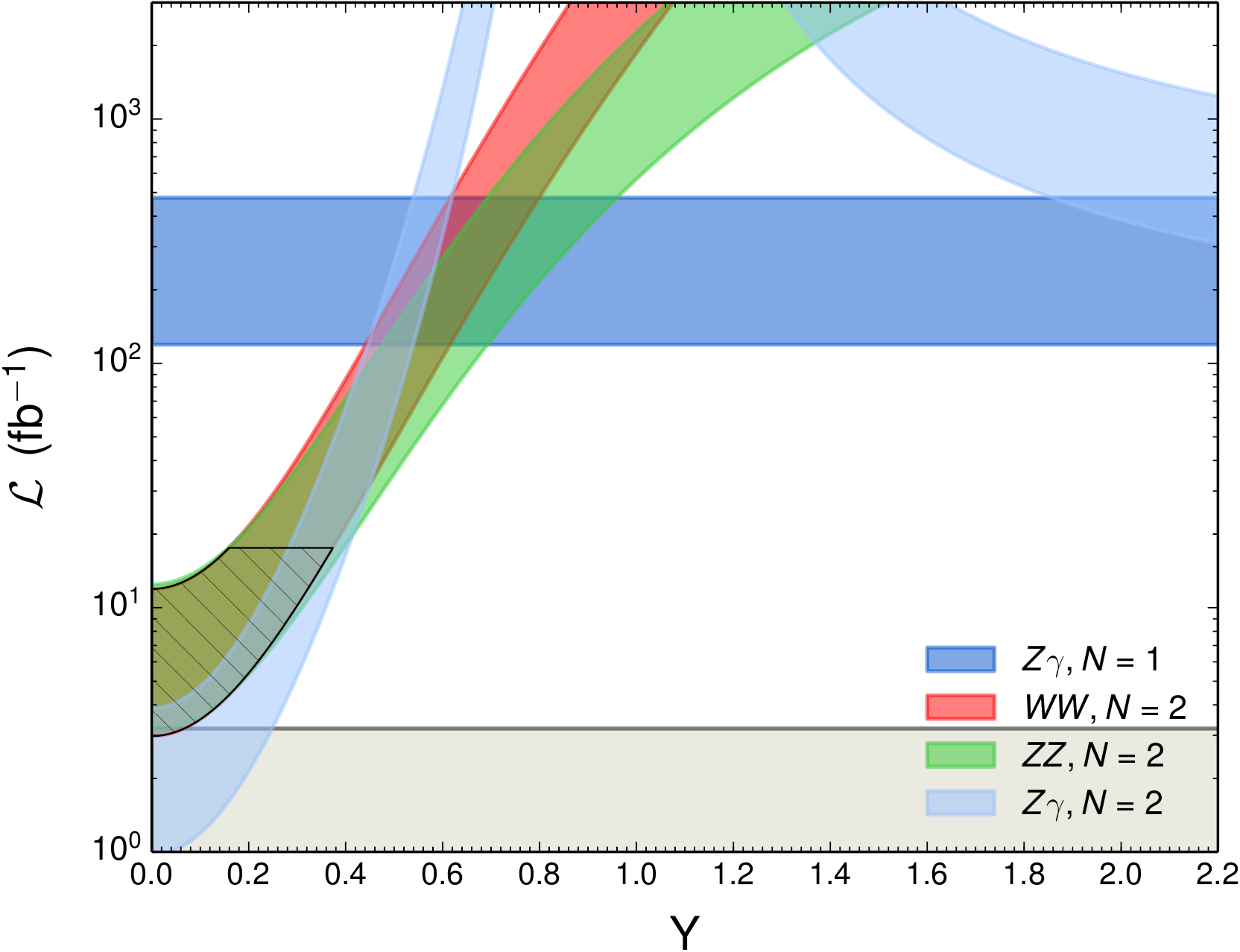}
\includegraphics[width=0.49\textwidth]{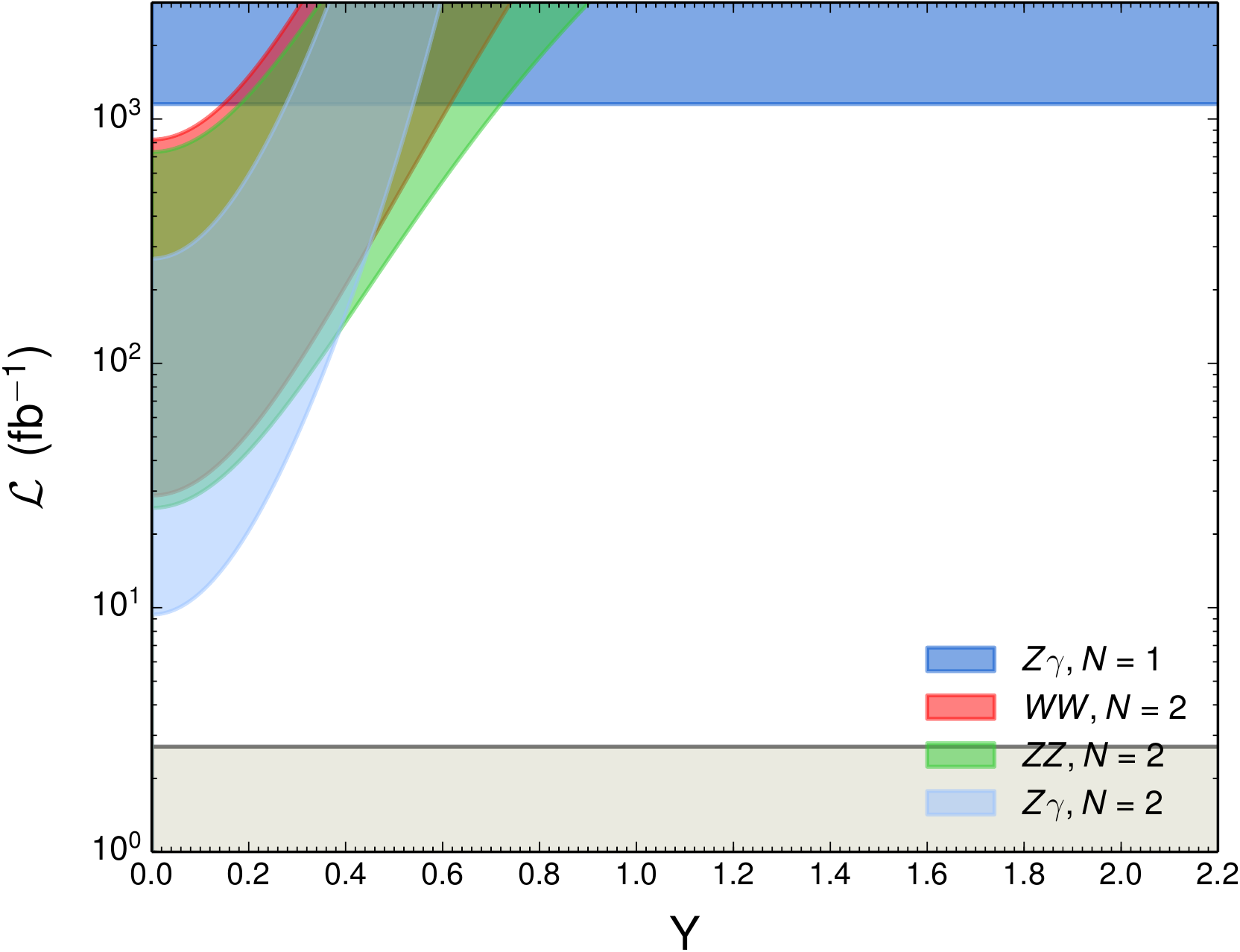}
\caption{Integrated luminosity $\mathcal{L}$ (in $\mathrm{fb}^{-1}$) needed to exclude at 95\% C.L. the presence of 
the decays $S\to Z\gamma$ (blue), $S\to ZZ$ (green) and $S\to WW$ (red), for an $SU(2)_{\mathrm{L}}$ singlet $\chi$ ($N = 1$) and 
$SU(2)_{\mathrm{L}}$ doublet $\chi$ ($N = 2$), as a function of the hypercharge Y of $\chi$ (for $N = 1$, Y $ > 0$ is implied). 
\textsl{Top}: NW scenario, assuming the di-photon production cross section $\sigma(p p \to S \to \gamma\gamma)$ range
favoured respectively by ATLAS (left), given in~\eqref{ATLAS_NW}, and by CMS (right), given in~\eqref{CMS_NW}.
\textsl{Bottom}: LW scenario, assuming the di-photon production cross section $\sigma(p p \to S \to \gamma\gamma)$ range
favoured respectively by ATLAS (left), given in~\eqref{ATLAS_LW}, and by CMS (right), given in~\eqref{CMS_LW}. Dashed areas show the exclusion regions from LHC 8 
TeV data (see text for details). The shaded light-grey region indicates the value of $\mathcal{L}$ from current 13 TeV searches.  
}
\label{fig2:Lumi}
\vspace{-4mm}
\end{figure}
\end{widetext}

The results are shown on Figure~\ref{fig2:Lumi}-\textsl{Top} for the case of NW assuming the di-photon production cross section ranges 
favoured respectively by ATLAS (left), given in~\eqref{ATLAS_NW}, and by CMS (right), given in~\eqref{CMS_NW}. 
Similarly, Figure~\ref{fig2:Lumi}-\textsl{Bottom} shows the results for the LW scenario, 
assuming respectively the di-photon production cross section ranges 
favoured  by ATLAS (left), given in~\eqref{ATLAS_LW}, and by CMS (right), given in~\eqref{CMS_LW}.

For $\chi$ being an $SU(2)_{\mathrm{L}}$ singlet ($N = 1$), the $W W$ decay mode is absent, while the 
$ZZ$ mode is beyond the LHC reach with end-of-lifetime integrated luminosity $\mathcal{L} = 3000\,\, \mathrm{fb}^{-1}$ both in the 
NW and LW scenarios. In contrast, the $Z\gamma$ decay mode would be within 13 TeV LHC reach with $\mathcal{L} = 100-500\,\, \mathrm{fb}^{-1}$ for 
the di-photon cross section range favoured by ATLAS (both for NW and LW). We however note that 
for the di-photon cross section range favoured by CMS, the required $\mathcal{L}$ to probe the $Z\gamma$ channel increases significantly.
This is particularly so for the LW scenario, as the local significance of the potential di-photon signal at $m_{\gamma\gamma} = 750$ GeV in the 13 TeV CMS data decreases from 
$N_{\sigma} \sim 2.87$ for NW to $N_{\sigma} \sim 2.52$ for the LW scenario~\cite{CMS:2016owr}. 
As a result of the above discussion, the $N = 1$ case may be excluded by probing the $WW$ and/or $ZZ$ decays modes of $S$ at the 13 TeV LHC, as well as by 
failing to observe the $S \to Z\gamma$ decay with $\mathcal{L} \sim \mathcal{O}(1000)\,\,\mathrm{fb}^{-1}$ (particularly if more precise measurements of the 
di-photon cross section agree with the range currently favoured by ATLAS data). 
%

For $N > 1$ and Y $> 0$, a di-photon cross section in the range favoured by ATLAS data would strongly 
suggest the existence of another gauge boson decay mode of $S$ within LHC reach, both for NW and LW, as shown in 
Figure~\ref{fig2:Lumi} (Left). These new channels would be $S \to Z Z$ and $S \to WW$ for $1 \lesssim \kappa \lesssim 6$ and an integrated luminosity 
$\mathcal{L} \gtrsim 30\,\,\mathrm{fb}^{-1}$, while for $\kappa \gtrsim 6$ the $S \to Z \gamma$ decay mode would be only one observed at LHC, which would at least need 
$\mathcal{L} \sim \mathcal{O}(100)\,\,\mathrm{fb}^{-1}$. For $\kappa \lesssim 1$, Figure~\ref{fig2:Lumi} shows that all three decay modes $S \to Z \gamma$,
$S \to Z Z$ and $S \to WW$ would be accesible, with $S \to Z \gamma$ most likely being the first channel to be discovered.

Finally, for $N > 1$ and Y $= 0$ ($\kappa = 0$), Figure~\ref{fig2:Lumi} shows that 13 TeV ATLAS and CMS $Z \gamma$ searches already strongly constrain this scenario with 
$\mathcal{L} \sim 3\,\,\mathrm{fb}^{-1}$, with the ATLAS and CMS favoured 13 TeV NW-CV di-photon cross sections being excluded at more than 95\% C.L., as well as the ATLAS 
favoured 13 TeV LW-CV di-photon cross section.

LHC 8 TeV ATLAS and CMS searches for spin-$0$ resonances decaying to $Z\gamma$~\cite{CMS:2016all}, $WW$~\cite{Aad:2015agg,Khachatryan:2015cwa} and 
$ZZ$~\cite{Aad:2015kna,Khachatryan:2015cwa} also yield strong constraints on small values of $\kappa$. As opposed to the 13 TeV constraints discussed above, in order to derive 
limits on $R_{VV}$ from 8 TeV data we need to assume a specific production mechanism for $S$ at the LHC, as the ratio 
$R_{\sigma} = \sigma_{13\,\mathrm{TeV}}(p p \to S)/\sigma_{8\,\mathrm{TeV}}(p p \to S)$ varies for different production mechanisms (being dependent on the parton 
luminosity evolution from 8 TeV to 13 TeV).
In the following we assume gluon fusion production of $S$, for which $R_{\sigma} \sim 4.7$ (for $m_S = 750$ GeV).
For a NW scenario, the 95\% C.L. limits from 8 TeV searches on the corresponding 13 TeV cross sections are 
$\sigma(p p \to S \to Z\gamma) < 25.5$ fb~\cite{CMS:2016all}, $\sigma(p p \to S \to WW) < 165$ fb~\cite{Aad:2015agg} and 
$\sigma(p p \to S \to ZZ) < 52$ fb~\cite{Aad:2015kna}. For a LW scenario, the only available analysis is that of~\cite{Aad:2015agg}, yielding
a 95\% C.L. limit on the corresponding 13 TeV cross section of $\sigma(p p \to S \to WW) < 201$ fb. All these limits are shown in 
Figure~\ref{fig2:Lumi} as dashed regions. As is apparent from Figure~\ref{fig2:Lumi}, the 8 TeV limits exclude the range of di-photon cross sections favoured by 13 TeV 
ATLAS data in the LW scenario at more than 95\% C.L. for $\kappa < 0.0085$. The ranges of di-photon cross sections favoured by 13 TeV 
ATLAS/CMS data in the NW scenario are also severely constrained. 
 
We note that a direct consequence of our analysis is that, would the 750 GeV resonance be observed in the $WW$ or $ZZ$ channels with an integrated luminosity 
$\mathcal{L} \lesssim 30 \,\,\mathrm{fb}^{-1}$ and without the prior/simultaneous observation of the corresponding $Z\gamma$ signature, such observation 
would not be possible to accommodate within the present analysis, which would indicate that $S$ is not a pure $SU(2)_{\mathrm{L}}$ singlet. 
We comment on this issue in the next Section.

\vspace{-3mm} 
 
\section{IV. Higgs-Singlet Mixing}
 
\vspace{-3mm}  
 
Let us now discuss the EW nature of $S$ itself, and its impact on the present analysis. First, we note that $S$ could be a neutral state from 
an $SU(2)_{\mathrm{L}}$ multiplet. However, if $S$ is part of a doublet, {\it e.g.}~as in a Two-Higgs-Doublet model, the alignment
limit~\cite{Gunion:2002zf} has to be invoked and even in this case several species of new vector-like fermion states $\chi$ with different 
$SU(2)_{\mathrm{L}} \times U(1)_{\mathrm{Y}}$ quantum numbers are needed (see {\it e.g.}~\cite{Altmannshofer:2015xfo}), leading to a rather 
barroque scenario. Accommodating the di-photon signature with higher $SU(2)_{\mathrm{L}}$ representations generically 
fails to achieve the required suppression of the tree-level decays 
$S\to WW,\,ZZ$~\cite{Chiang:2016ydx} to comply with bounds from ATLAS and CMS searches at LHC 8 
TeV. 
The above issues then favour $S$ being predominantly an EW singlet. If it is a pseudoscalar, then mixing of $S$ with the SM Higgs is automatically avoided 
and the analysis carried out in this work fully applies. On the other hand, if $S$ is a scalar, then mixing of $S$ with the SM Higgs will 
occur (even if not present at tree-level, it will always occur at some loop order), leading to tree-level decays $S\to WW,\,ZZ$ which could potentially 
modify the results of the present analysis. We now quantify the range of validity of our analysis in the presence of 
Higgs-singlet mixing, parametrized here by 
$\mathrm{sin}\, \theta \equiv s_{\theta}$, 
and then discuss the experimental constraints on the value of such mixing. 

%
%

Let us consider for simplicity $V V = W W$, and $\chi$ to be an $n$-tuple of vector-like 
fermions, whose Yukawa-type coupling to $S$ reads   
\begin{equation}
\label{yukawa}
\lambda_{S\chi} S\, \bar{\chi} \chi \,.
\end{equation}
Our analysis from Sections~II and~III is strictly valid in the limit $s_{\theta} \to 0$. In this limit, the partial decay width $\Gamma_{WW}$ induced at 
1-loop by~\eqref{yukawa} (implicitly assuming $N > 1$) is given by (see {\it e.g.}~\cite{Altmannshofer:2015xfo})
\begin{eqnarray} 
\Gamma^{\mathrm{loop}}_{WW} &\simeq& \frac{2\,\alpha^2_{\mathrm{EM}}\,m_S}{144^2\,\pi^3\,s^4_{W}}\,n^2\,\lambda^2_{S\chi} \mathcal{A}_{\frac{1}{2}}^2(\tau_{\chi}) \nonumber \\
&\times& \left[N(N-1)(N+1) \right]^2
\end{eqnarray}
with $n$ also accounting for potential colour degrees of freedom, $\tau_{\chi} = m_S^2/(4\,m^2_{\chi}) < 1$, and $A_{\frac{1}{2}}(x)$ being a loop function which can be found 
{\it e.g.}~in~\cite{Carmi:2012in}.

In the presence of Higgs-singlet mixing $s_{\theta} \neq 0$, but now in the absence of new states $\chi$ which would yield loop-induced contributions to $S \to WW$, the decay 
is entirely due to Higgs-singlet mixing, with a partial decay width given by~\cite{Dittmaier:2011ti}
\begin{equation}
\label{gammaWtree}
\Gamma^{\mathrm{tree}}_{WW} = s^2_{\theta}\times \Gamma^{\mathrm{SM}}_{WW}(m_S) = s^2_{\theta}\times 145 \,\,\mathrm{GeV}\,.
\end{equation}
The consistency of our analysis requires $\Gamma^{\mathrm{tree}}_{WW}/\Gamma^{\mathrm{loop}}_{WW} \lesssim 1$, which leads to
\begin{equation}
\label{thetabound}
|s_{\theta}| \lesssim 1.3\times10^{-4}\left[n\lambda_{S\chi} \mathcal{A}_{\frac{1}{2}}(\tau_{\chi})N(N-1)(N+1) \right].
\end{equation}
A similar bound may be obtained (now involving the value of the hypercharge Y) for $VV = Z Z$.
Since models fitting the di-photon excess typically require the bracketed term in~\eqref{thetabound} to be 
$\sim \mathcal{O}(100)$~\cite{No:2015bsn}, this yields an approximate bound $|s_{\theta}| \lesssim 10^{-2}$.

The above discussion implies that for a pseudoscalar $S$ our results are generically robust. For a scalar $S$, our analysis fully applies if the 
Higgs-singlet mixing occurs at 1-loop or higher (since then $|s_{\theta}| < (16\pi^2)^{-1}$ is expected) whereas for tree-level Higgs-singlet mixing a 
high degree of tuning is required. This also highlights the fact that an SM$+S$ EFT analysis of the di-photon anomaly, 
with $S$ being an $SU(2)_{\mathrm{L}} \times U(1)_{\mathrm{Y}}$ singlet scalar, requires the inclusion of Higgs-singlet 
mixing effects~\cite{Franceschini:2016gxv,Gripaios:2016xuo,Dawson:2016ugw}.

\vspace{3mm}

Turning to the analysis of experimental bounds on the mixing, we first note that the amount 
of mixing $s_{\theta}$ is severely constrained by measurements of EW precision observables (EWPO). 
We can safely neglect the impact of the new states $\chi$ on EWPO for $m_{\chi} > 375$ GeV (as the region of parameter space where they could be relevant 
is ruled out by Drell-Yan measurements at LHC 8 TeV~\cite{Gross:2016ioi}), 
and then EWPO directly constrain the presence of Higgs-singlet mixing. The shift on any 
oblique parameter $\mathcal{O} = \mathrm{S},\, \mathrm{T}, \, \mathrm{U}$ 
due to the Higgs-singlet mixing can be written entirely in terms of the SM Higgs contribution to that 
parameter, $\mathcal{O}^{\mathrm{SM}}(m)$, where $m$ is either $m_h$ or $m_S$
\begin{eqnarray}
& \Delta \mathcal{O} = (c^2_{\theta} -1) \mathcal{O}^{\mathrm{SM}}(m_h) + s^2_{\theta} \; \mathcal{O}^{\mathrm{SM}}(m_S) & \nonumber \\
& = s^2_{\theta} \left( \mathcal{O}^{\mathrm{SM}}(m_S) - \mathcal{O}^{\mathrm{SM}}(m_h) \right) ,&
\end{eqnarray}
Taking the best-fit values for the shifts $\Delta \mathcal{O}$ from the latest EW fit to the SM by the 
Gfitter group~\cite{Baak:2014ora} and performing a $\chi^2$ fit to these data (for details, see~\cite{Gorbahn:2015gxa}) leads to the bound 
$|s_{\theta}| < 0.24$.

Another powerful source of constraints on the value of $s_{\theta}$ comes from LHC 8 TeV ATLAS and CMS searches for high-mass spin-$0$ resonances decaying 
to $WW$ and $ZZ$~\cite{Aad:2015agg,Aad:2015kna,Khachatryan:2015cwa}. We note that these limits, discussed in the previous Section, 
depend mildly on the amount of mixing (as it influences the width of $S$), as well as on the LHC production mechanism for $S$ and the 
choice of favoured range for the 13 TeV di-photon cross section (for ATLAS or CMS). 
We consider gluon fusion production for $S$, and neglect for simplicity the 1-loop contributions to the decay amplitudes for 
$S \to WW$ and $S\to ZZ$, in order to disregard the possible effect of interplay between the tree-level 
and 1-loop contributions to the amplitudes (see~\cite{Dawson:2016ugw} for a recent discussion of these effects). 
We can then give an estimate of the upper bound on $\Gamma^{\mathrm{tree}}_{VV}/\Gamma_{\gamma\gamma}$ as a bound on $R_{VV}$ 
from 8 TeV searches (see Figure~\ref{fig2:Lumi}). For $V = W$, $\Gamma^{\mathrm{tree}}_{WW}$ is given by~\eqref{gammaWtree}, while for 
$V = Z$, $\Gamma^{\mathrm{tree}}_{ZZ} = s^2_{\theta}\times 72$ GeV~\cite{Dittmaier:2011ti}. The di-photon partial width of $S$ is given by
\begin{eqnarray} 
\Gamma_{\gamma\gamma} &=& \frac{\alpha^2_{\mathrm{EM}}\,m_S}{144^2\,\pi^3}\,n^2\,\lambda^2_{S\chi} \mathcal{A}_{\frac{1}{2}}^2(\tau_{\chi}) \nonumber \\
&\times& N^2\left[12\,\mathrm{Y}^2+(N-1)(N+1) \right]^2\, .
\end{eqnarray}
Bearing in mind that fitting the di-photon excess requires 
$n \lambda_{S\chi} \mathcal{A}_{\frac{1}{2}}(\tau_{\chi}) N \left[12\,\mathrm{Y}^2+(N-1)(N+1) \right] \sim \mathcal{O}(100)$, and 
considering for a conservative estimate the lower edge of the ATLAS and CMS favoured ranges for the di-photon signal 
cross section, we obtain a rough bound $|s_{\theta}| \lesssim 0.03$, in good agreement with~\cite{Dawson:2016ugw}.
The fact that this bound is of the same order as the bound on $|s_{\theta}|$ for the consistency of our analysis, given by~\eqref{thetabound}, 
highlights that current limits on $R_{VV}$ from 8 TeV data are of similar order as the values of $R_{VV}$ for a pure 
$SU(2)_{\mathrm{L}}$ singlet $S$ decaying to SM gauge bosons via loops of $\chi$ particles\footnote[4]{This also stresses the importance of 
accounting for the interplay between tree-level and 1-loop contributions to the decay amplitude, in the presence of Higgs-singlet mixing, for a precise 
bound on $s_{\theta}$~\cite{Dawson:2016ugw}.}
with $SU(2)_{\mathrm{L}} \times U(1)_{\mathrm{Y}}$ quantum numbers such that $\kappa \ll 1$, as discussed in Section II.

\vspace{-3mm} 
 
\section{V. conclusions}
 
\vspace{-3mm}   
 
The existence of a $750$ GeV di-photon spin-$0$ resonance $S$, if confirmed by upcoming LHC data, would have profound 
implications for physics beyond the SM. Among them would be the additional presence of new degrees of freedom $\chi$ coupled to 
$S$ and electromagnetically charged. Unravelling the properties of these new states would them become a key task for LHC. In this work 
I have discussed a powerful probe of the $SU(2)_{\mathrm{L}} \times U(1)_{\mathrm{Y}}$ quantum numbers of these new states $\chi$, through 
measurements of the ratios $R_{VV}$ between the cross sections $pp \to S\to VV$ with $VV=Z\gamma,\,ZZ,\,WW$ and the di-photon cross section measured 
by ATLAS and CMS. These probes are independent of the bosonic/fermionic nature of the new states $\chi$, their mass and 
coupling to $S$. I have provided robust estimates of the amount of LHC integrated luminosity $\mathcal{L}$ required to explore various possible 
scenarios, showing in particular that LHC has the capability to disentangle the pure $SU(2)_{\mathrm{L}}$ case, the pure $U(1)_{\mathrm{Y}}$ case and the 
case when $\chi$ transform under both symmetries, from each other. This would provide 
very useful information on the underlying theory beyond the SM, as well as on the strategy to search for these new states directly at the LHC and beyond. 
as well as on the strategy to search for these new states directly at the LHC and beyond. 

\begin{center}
\textbf{Acknowledgements} 
\end{center}

\vspace{-2mm}

I want to thank Tilman Plehn, Veronica Sanz and Ken Mimasu for useful discussions and comments on the manuscript, 
as well as the Mitchell Institute for Theoretical Physics from Texas A\&M University and the 
Institute for Theoretical Physics from Heidelberg University, for their kind hospitality while this work was being performed.
J.M.N. is supported by the People Programme (Marie Curie Actions) of the European Union Seventh 
Framework Programme (FP7/2007-2013) under REA grant agreement PIEF-GA-2013-625809.

\end{document}